\def\papertitle{The Shape of RemiXXXes to Come: Audio Texture Synthesis with Time--frequency Scattering}
\def\paperauthorA{Vincent Lostanlen}
\def\paperauthorB{Florian Hecker}
\def\paperauthorC{Author Three}
\def\paperauthorD{Author Four}

\documentclass[twoside,a4paper]{article}
\usepackage{dafx_19}
\usepackage{amsmath,amssymb,amsfonts,amsthm}
\usepackage{euscript}
\usepackage[latin1]{inputenc}
\usepackage[T1]{fontenc}
\usepackage{ifpdf}
\usepackage{siunitx}
\usepackage{stmaryrd}
\usepackage{url}
\usepackage[english]{babel}
\usepackage{color}
\usepackage{xspace}
\usepackage{subcaption}
\usepackage{microtype}

\newcommand\blfootnote[1]{%
  \begingroup
  \renewcommand\thefootnote{}\footnote{#1}%
  \addtocounter{footnote}{-1}%
  \endgroup
}

\newcommand*{\eg}{e.g.\@\xspace}
\newcommand*{\ie}{i.e.\@\xspace}
\newcommand*{\resp}{resp.\@\xspace}

\ninept
\usepackage{times}

\newif\ifpdf
\ifx\pdfoutput\relax
\else
   \ifcase\pdfoutput
      \pdffalse
   \else
      \pdftrue
\fi

\ifpdf 
  \usepackage[pdftex,
    pdftitle={\papertitle},
    pdfauthor={\paperauthorA, \paperauthorB, \paperauthorC, \paperauthorD},
    colorlinks=false, 
    bookmarksnumbered, 
    pdfstartview=XYZ 
  ]{hyperref}
  \usepackage[pdftex]{graphicx}
  \usepackage[figure,table]{hypcap}
\else 
  \usepackage[dvips]{epsfig,graphicx}
  \usepackage[dvips,
    colorlinks=false, 
    bookmarksnumbered, 
    pdfstartview=XYZ 
  ]{hyperref}
  \usepackage[figure,table]{hypcap}
\fi

\twoaffiliations{\paperauthorA}
{\href{https://steinhardt.nyu.edu/marl/}{Music and Audio Research Lab} \\
Steinhardt School of Culture, Education, and Human Development \\
New York University \\ 
New York, NY, USA \\
{\tt \href{mailto:vincent.lostanlen@nyu.edu}{vincent.lostanlen@nyu.edu}}
}
{\paperauthorB}
{\href{https://www.eca.ed.ac.uk/profile/florian-hecker}{Edinburgh College of Art} \\ 
College of Arts, Humanities, and Social Sciences \\
University of Edinburgh \\
Edinburgh, UK \\
{\tt \href{mailto:florian.hecker@ed.ac.uk}{florian.hecker@ed.ac.uk}}
}
\begin{document}
\ifpdf 
  \DeclareGraphicsExtensions{.png,.jpg,.pdf}
\else  
  \DeclareGraphicsExtensions{.pdf}
\fi

\capstartfalse
\maketitle

\begin{abstract}
This article explains how to apply time--frequency scattering, a convolutional operator extracting modulations in the time--frequency domain at different rates and scales, to the re-synthesis and manipulation of audio textures.
After implementing phase retrieval in the scattering network by gradient backpropagation, we introduce \emph{scale-rate DAFx}, a class of audio transformations expressed in the domain of time--frequency scattering coefficients.
One example of scale-rate DAFx is chirp rate inversion, which causes each sonic event to be locally reversed in time while leaving the arrow of time globally unchanged.
Over the past two years, our work has led to the creation of four electroacoustic pieces: \emph{FAVN}; \emph{Modulator (Scattering Transform)}; \emph{Experimental Palimpsest}; \emph{Inspection (Maida Vale Project)} and \emph {Inspection II}; as well as \emph{ XAllegroX (Hecker Scattering.m Sequence)},  a remix of Lorenzo Senni's \emph{XAllegroX}, released by Warp Records on a vinyl entitled \emph{The Shape of RemiXXXes to Come}.\blfootnote{This work is supported by the ERC InvariantClass grant 320959. The source code to reproduce experiments and figures is made freely available at: \url{https://github.com/lostanlen/scattering.m}}
\end{abstract}
\capstarttrue

\section{Introduction}
\label{sec:intro}
Several composers have pointed out the lack of a satisfying trade-off between interpretability and flexibility in the parametrization of sound transformations \cite{kaper1999icmc,risset2007smc, cella2017dlm}.
For example, the constant-$Q$ wavelet transform (CQT) of an audio signal provides an intuitive display of its short-term energy distribution in time and frequency  \cite{velasco2011dafx}, but does not give explicit control over its intermittent perceptual features, such as roughness or vibrato.
On the other hand, a deep convolutional generative model such as WaveNet \cite{engel2017icml} encompasses a rich diversity of timbre; but, because the mutual dependencies between the dimensions of its latent space are unspecified, music composition with autoencoders in the waveform domain is hampered by a long preliminary phase of trials and errors in the search for the intended effect.

Scattering transforms are a class of multivariable signal representations
at the crossroads between wavelets and deep convolutional networks \cite{mallat2016philosophical}.
In this paper, we demonstrate that one such instance of scattering transform, namely time--frequency scattering \cite{anden2015mlsp}, can be a relevant tool for composers of electroacoustic music, as it strikes a satisfying compromise between interpretability and flexibility.
We describe the scattering-based DAFx underlying the synthesis of five electroacoustic pieces by Florian Hecker: \emph{FAVN} (2016); \emph{Modulator (Scattering Transform)} (2016-2017); \emph{Experimental Palimpsest} (2016); \emph{Inspection (Maida Vale Project)}(2016) and \emph{Inspection II} (2017); as well as \emph{ XAllegroX (Hecker Scattering.m Sequence)}, a remix of Lorenzo Senni's \emph{XAllegroX}, released by Warp Records on a vinyl entitled \emph{The Shape of RemiXXXes to Come} (2017).
In addition, we demonstrate the result of this algorithm in the companion website of this paper, which contains short audio examples as well as links to full-length computer-generated sound pieces.
\blfootnote{Companion website: \url{https://lostanlen.com/pubs/dafx2019}}

Section 2 defines time--frequency scattering.
Section 3 presents a gradient backpropagation method for sound synthesis from time--frequency scattering coefficients.
Section 4 introduces ``scale-rate DAFx'', a new class of DAFx which operates in the domain of spectrotemporal modulations, and describes the implementation of chirp reversal as a proof of concept.

\section{Time--frequency scattering}
\label{sec:time--frequency-scattering}

\begin{figure}
\centerline{\includegraphics[height=3cm]{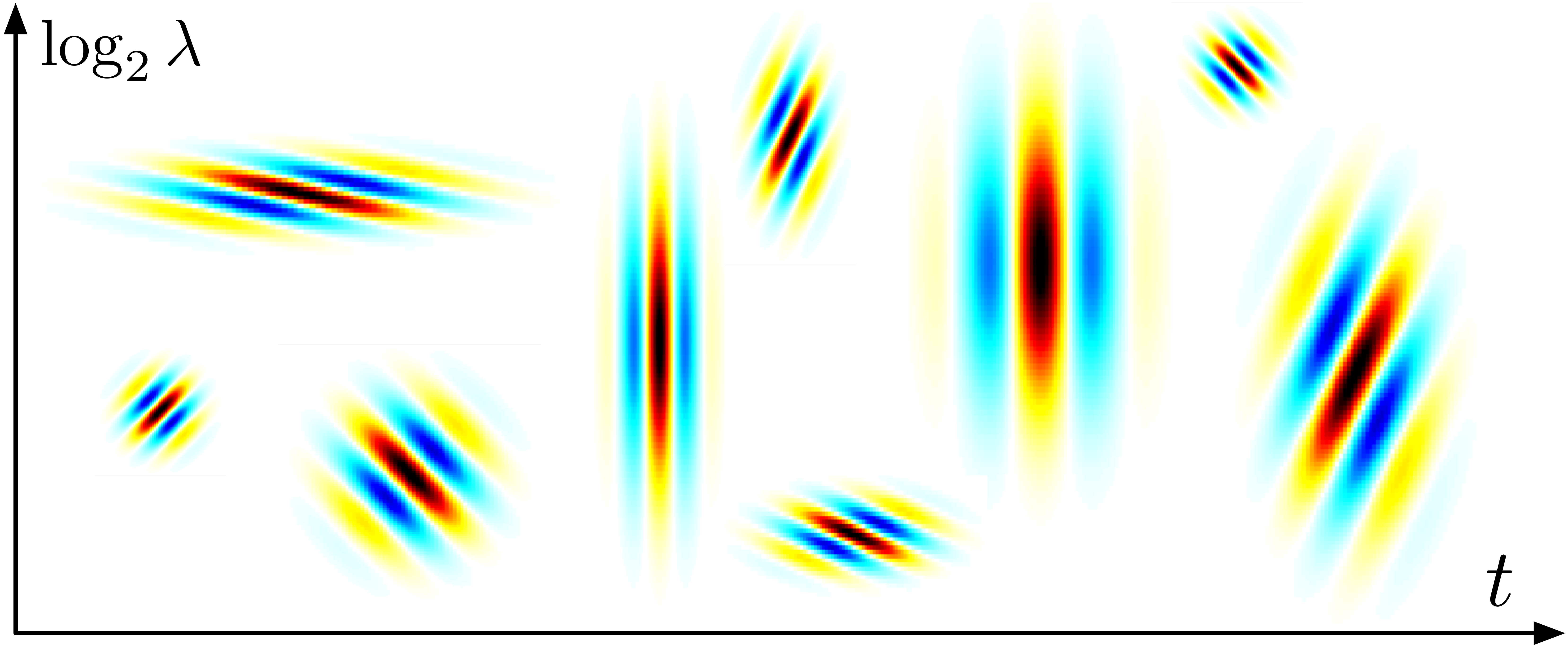}}
\caption{Interference pattern between wavelets $\boldsymbol{\psi}_\alpha (t)$ and $\boldsymbol{\psi}_\beta (\log_2 \lambda)$ in the time--frequency domain $(t, \log_2 \lambda)$ for different combinations of amplitude modulation rate $\alpha$ and frequency modulation scale $\beta$. Darker shades of red (resp.~blue) indicate higher positive (resp.~lower negative) values of the real part. See Section 2 for details.}
\label{fig:spectrotemporal-wavelets}
\end{figure}

In this section, we define the time--frequency scattering transform as a function of four variables --- time $t$, frequency $\lambda$, amplitude modulation rate $\alpha$, and frequency modulation scale $\beta$ --- which we connect to spectrotemporal receptive fields (STRF) in auditory neurophysiology \cite{patil2012plos}.
We refer to \cite{anden2019tsp} for an in-depth mathematical introduction to time--frequency scattering.

\subsection{Spectrotemporal receptive fields}
Time--frequency scattering results from the cascade of two stages: a constant-$Q$ wavelet transform (CQT) and the extraction of spectrotemporal modulations with wavelets in time and log-frequency.
First, we define Morlet wavelets of center frequency $\lambda > 0$ and quality factor $Q$ as
\begin{equation}
\boldsymbol{\psi}_{\lambda}(t) =
\lambda
\exp\left(- \dfrac{\lambda^2 t^2}{2Q^2}\right) \times
( \exp(2\pi \mathrm{i} \lambda t) - \kappa),
\end{equation}
where the corrective term $\kappa$ ensures that each $\boldsymbol{\psi}_\lambda (t)$ has one vanishing moment, \ie a null average.
In the sequel, we set $Q=12$ to match twelve-tone equal temperament.
Within a discrete setting, acoustic frequencies $\lambda$ are typically of the form $2^{n/Q}$ where $n$ is integer, thus covering the hearing range.
For $\boldsymbol{x}(t)$ a finite-energy signal, we define the CQT of $\boldsymbol{x}$ as the matrix
\begin{equation}
\mathbf{U}_1 \boldsymbol{x}(t, \lambda) =
\left \vert \boldsymbol{x} \ast \boldsymbol{\psi}_{\lambda} \right \vert (t),
\label{eq:U1}
\end{equation}
that is, stacked convolutions with all wavelets $\boldsymbol{\psi}_\lambda (t)$ followed by the complex modulus nonlinearity.

Secondly, we define Morlet wavelets of respective center frequencies $\alpha > 0$ and $\beta \in \mathbb{R}$ with quality factor $Q=1$.
With a slight abuse of notation, we denote these wavelets by $\boldsymbol{\psi}_\alpha(t)$ and $\boldsymbol{\psi}_\beta(\log \lambda)$ even though they do not necessarily have the same shape as the wavelets $\boldsymbol{\psi}_{\lambda}(t)$ of Equation \ref{eq:U1}. Frequencies $\alpha$, hereafter called amplitude modulation \emph{rates}, are measured in Hertz (Hz) and discretized as $2^n$ with integer $n$.
Frequencies $\beta$, hereafter called frequency modulation \emph{scales}, are measured in cycles per octave (c/o) and discretized as $\pm 2^n$ with integer $n$.
The edge case $\beta = 0$ corresponds to $\boldsymbol{\psi}_{\beta}(\log \lambda)$ being a Gaussian low-pass filter $\boldsymbol{\phi}_F (\log \lambda)$ of bandwidth $F^{-1}$.
These modulation scales $\beta$ play the same role as the \emph{quefrencies} in a mel-frequency cepstrum \cite{anden2015mlsp}.

We define the spectrotemporal receptive field (STRF) of $\boldsymbol{x}$ as the fourth-order tensor
\begin{multline}
\mathbf{U}_2 \boldsymbol{x}(t, \lambda, \alpha, \beta) =
\big \vert \mathbf{U}_1 \boldsymbol{x} \overset{t}{\ast} \boldsymbol{\psi}_{\alpha} \overset{\log_2 \lambda}{\ast}  \boldsymbol{\psi}_{\beta} \big \vert (t, \lambda) \\
=
\Bigg\vert\iint\! \mathbf{U}_1 \boldsymbol{x} (\tau, s) \boldsymbol{\psi}_{\alpha}(t-\tau) \boldsymbol{\psi}_{\beta}(\log_2 \lambda - s) \,\mathrm{d}\tau \,\mathrm{d}s\Bigg\vert,
\label{eq:U2}
\end{multline}
that is, stacked convolutions in time and log-frequency with all wavelets $\boldsymbol{\psi}_\alpha (t)$ and $\boldsymbol{\psi}_\beta (\log_2 \lambda)$ followed by the complex modulus nonlinearity \cite{lindeberg2015plos}. 
Figure \ref{fig:spectrotemporal-wavelets} shows the interference pattern of the product $\boldsymbol{\psi}_{\alpha}(t - \tau) \boldsymbol{\psi}_\beta (\log_2 \lambda - s)$ for different combinations of time $t$, frequency $\lambda$, rate $\alpha$, and scale $\beta$.
We denote the multiindices $(\lambda, \alpha, \beta)$ resulting from such combinations as scattering \emph{paths} \cite{mallat2012cpam}.
We refer to \cite{siedenburg2016jnmr} for an introduction to STRFs in the interdisciplinary context of music cognition and music information retrieval (MIR), and to \cite{schadler2012jasa} for an experimental benchmark in automatic speech recognition.

\subsection{Invariance to translation}
Because it is a convolutional operator in the time--frequency domain, the STRF is equivariant to temporal translation $t \mapsto t + \tau$ as well as frequency transposition $\lambda \mapsto 2^s \lambda$.
In audio classification, it is useful to guarantee invariance to temporal translation up to some time lag $T$ \cite{mallat2012cpam}.
To this aim, we define time--frequency scattering as the result of a local averaging of both $\mathbf{U_1} \boldsymbol{x} (t, \lambda)$ and $\mathbf{U}_2 \boldsymbol{x}(t, \lambda, \alpha, \beta)$ by a Gaussian low-pass filter $\boldsymbol{\phi}_T$ of cutoff frequency equal to $T^{-1}$, yielding
\begin{equation}
\mathbf{S}_1 \boldsymbol{x}(t, \lambda) =
\big( \mathbf{U}_1 \boldsymbol{x} \overset{t}{\ast} \boldsymbol{\phi}_T \big)(t, \lambda) \textrm{\quad and}
\label{eq:S1}
\end{equation}
\begin{equation}
\mathbf{S}_2 \boldsymbol{x}(t, \lambda, \alpha, \beta) =
\big( \mathbf{U}_2 \boldsymbol{x}\overset{t}{\ast} \boldsymbol{\phi}_T \big)(t, \lambda, \alpha, \beta)
\label{eq:S2}
\end{equation}
respectively.
In practice, for purposes of signal classification, $T$ is of the order of \SI{50}{\milli\second} in speech; of \SI{500}{\milli\second} in instrumental music; and of \SI{5}{\second} in ecoacoustics \cite{lostanlen2017phd}.
The delay of a real-time implementation of time--frequency scattering is of the order of $T$.

\subsection{Energy conservation}
\label{sub:energy-conservation}
\begin{figure}
\centering
   \begin{subfigure}[b]{\linewidth}
   \includegraphics[width=\linewidth]{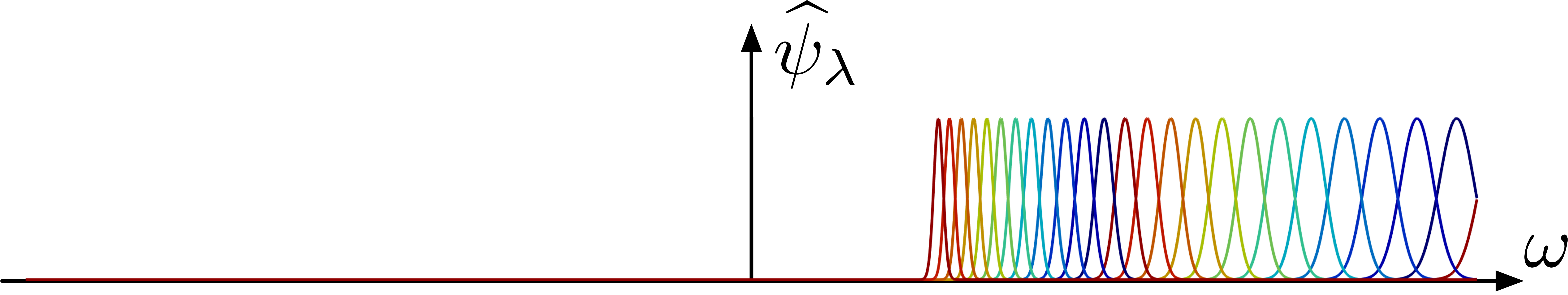}
   \caption{Wavelets $\widehat{\boldsymbol{\psi}}_\lambda$ of frequencies $\lambda$ and quality factor $Q = 12$.}
   \label{fig:energy-conservation-a} 
\end{subfigure}
\begin{subfigure}[b]{\linewidth}
   \includegraphics[width=\linewidth]{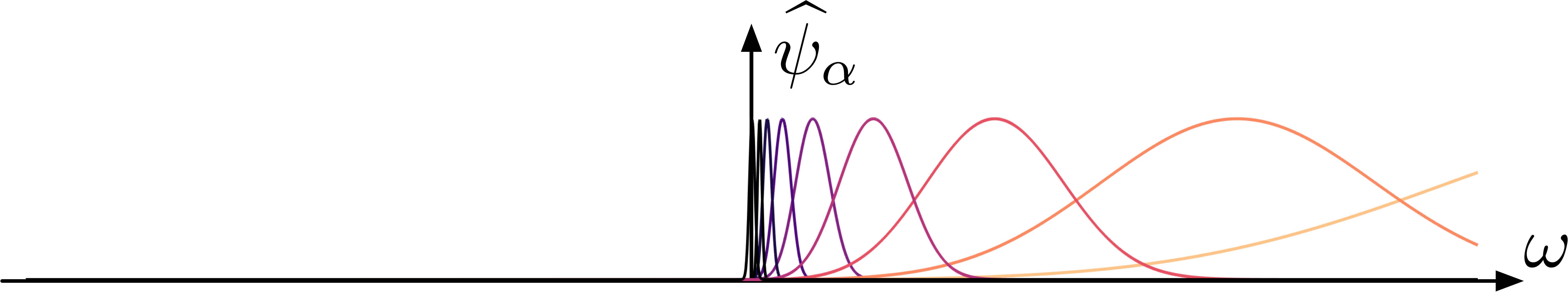}
   \caption{Wavelets $\widehat{\boldsymbol{\psi}}_\alpha$ of rates $\alpha$ and quality factor $Q = 1$.}
   \label{fig:energy-conservation-b}
\end{subfigure}
\begin{subfigure}[b]{\linewidth}
   \includegraphics[width=\linewidth]{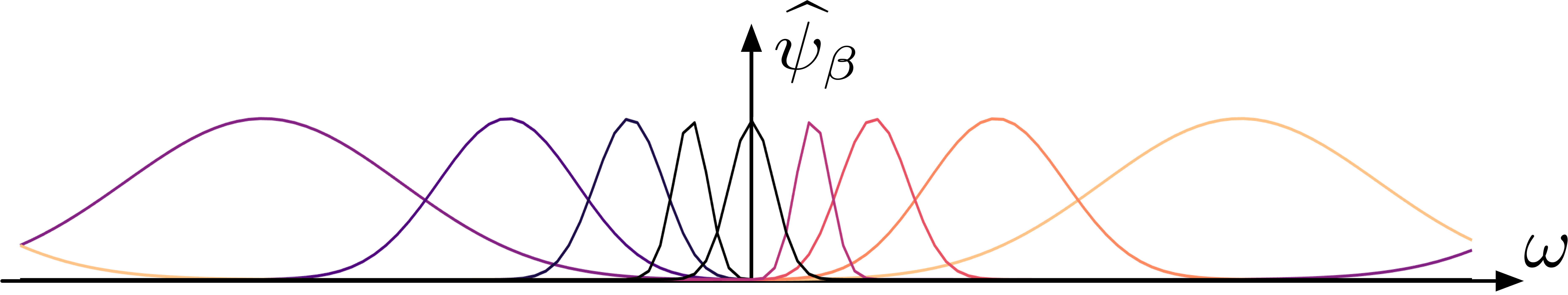}
   \caption{Wavelets $\widehat{\boldsymbol{\psi}}_\beta$ of scales $\beta$ and quality factor $Q = 1$.}
   \label{fig:energy-conservation-c}
\end{subfigure}
\caption{Filterbanks of Morlet wavelets in the Fourier domain:
(a) for CQT;
(b) for STRF, temporal dimension;
(c) for STRF, log-frequential dimension. See Section 2 for details.}
\label{fig:energy-conservation}
\end{figure}

We restrict the set of modulation rates $\alpha$ in $\mathbf{U_2}\boldsymbol{x}$ to values above $T^{-1}$, so that the power spectra of the low-pass filter $\boldsymbol{\phi}_T (t)$ and all wavelets $\boldsymbol{\psi}_\alpha (t)$ cover uniformly the Fourier domain \cite[Chapter~4]{mallat2008book}:
at every frequency $\omega$, we have
\begin{equation}
\big\vert \boldsymbol{\widehat{\phi}}_T (\omega)\big\vert^2 + \frac{1}{2} \sum_{\alpha > T^{-1}} \Big( \big\vert \boldsymbol{\widehat{\psi}}_\alpha (\omega)\big\vert^2 + \big\vert \boldsymbol{\widehat{\psi}}_\alpha (-\omega)\big\vert^2\Big) \lessapprox 1,
\end{equation}
where the notation $A \lessapprox B$
indicates that there exists some $\varepsilon \ll B$ such that $B - \varepsilon < A < B$.
In the Fourier domain associated to $\log_2 \lambda$, one has $ \sum_{\beta} \vert \boldsymbol{\widehat{\psi}}_\beta (\omega)\vert^2 \lessapprox 1$ for all $\omega$.
Therefore, applying Parseval's theorem on all three wavelet filterbanks (respectively indexed by $\lambda$, $\alpha$, and $\beta$) yields
$\Vert \mathbf{S}_1 \boldsymbol{x} \Vert_2^2  + \Vert \mathbf{U}_2 \boldsymbol{x} \Vert_2^2 \lessapprox
\Vert \mathbf{U}_1 \boldsymbol{x} \Vert_2^2$.
Figure \ref{fig:energy-conservation} illustrates the design of these filterbanks in the Fourier domain.

The spectrotemporal modulations in music --- \eg{} tremolo, vibrato, and dissonance --- are captured and demodulated by the second layer of a scattering network \cite{anden2012dafx}.
Consequently, each scattering path $(\lambda, \alpha, \beta)$ in $\mathbf{U}_2 \boldsymbol{x}(t, \lambda, \alpha, \beta)$ yields a time series whose variations are slower than in the first layer $\mathbf{U}_1 \boldsymbol{x}(t, \lambda)$; typically at rates of \SI{1}{\Hz} or lower.
By setting $T$ to $1$ second or less, we may safely assume that the cutoff frequency of the low-pass filter $\boldsymbol{\phi}_T(t)$ in Equation \ref{eq:S2} is high enough to retain all the energy in $\mathbf{U}_2 \boldsymbol{x}(t, \lambda, \alpha, \beta)$.
This assumption writes as $\Vert \mathbf{S}_2 \boldsymbol{x} \Vert \lessapprox \Vert \mathbf{U}_2 \boldsymbol{x} \Vert$ and is justified by the theorem of exponential decay of scattering coefficients \cite{waldspurger2017sampta}.
Let $\mathbf{S}$ be the operator resulting from the concatenation of first-order scattering $\mathbf{S_1}$ with second-order scattering $\mathbf{S_2}$.
In the absence of any DC bias in $\boldsymbol{x}(t)$, we conclude with the energy conservation identity
\begin{equation}
\Vert \mathbf{S} \boldsymbol{x} \Vert_2^2 =
\Vert \mathbf{S}_1 \boldsymbol{x} \Vert_2^2 +
\Vert \mathbf{S}_2 \boldsymbol{x} \Vert_2^2 \lessapprox
\Vert \mathbf{U}_1 \boldsymbol{x} \Vert_2^2 \lessapprox
\Vert \boldsymbol{x} \Vert_2^2.
\end{equation}

\section{Audio texture synthesis}
\label{sec:audio-texture-synthesis}

In this section, we describe how to pseudo-invert time--frequency scattering, that is, to generate a waveform whose scattering coefficients match the scattering coefficients of some other, pre-recorded waveform.

\subsection{From phase retrieval to texture synthesis}
\label{sub:phase-retrieval}
Although the invertibility of the convolutional operator involved in the constant-$Q$ transform is guaranteed by wavelet frame theory \cite[Chapter~5]{mallat2008book}, the complex modulus nonlinearity in Equation \ref{eq:U1} raises a fundamental question:
is it always possible to recover $\boldsymbol{x}$, up to a constant and therefore imperceptible phase shift, from the magnitudes of its wavelet coefficients $\mathbf{U}_1 \boldsymbol{x} (t,\lambda)$ ?
This question has recently been answered in the affirmative, suggesting that a redundant CQT should be preferred over critically sampled short-term Fourier transforms (STFT) when attempting to sonify spectrograms in the absence of any prior information on the phase of the target waveform \cite{waldspurger2017tit}.

By recursion over layers in the composition of wavelet modulus operators, the invertibility of wavelet modulus implies the invertibility of scattering transforms of infinite depth \cite{waldspurger2015phd}, up to a constant time shift of at most $T$.
However, the time--frequency scattering network presented here has a finite number of layers (\ie{}, most usually two layers) and is therefore not exactly invertible.
Because of the theorem of exponential decay of scattering coefficients \cite{waldspurger2017sampta}, the residual energy that is present in deeper layers can be neglected on the condition that $T$ is small enough.
In the rest of this section, we set $T$ to $\SI{186}{\milli\second}$, which corresponds to $8192$ samples at a sample rate of \SI{44.1}{\kilo\hertz}.
Since the unit circle is not a convex subset of $\mathbf{R}^2$, the optimization problem
$\boldsymbol{y}^\ast (t) =  \arg \min_{\boldsymbol{y}} \mathbf{E}(\boldsymbol{y})$
where
$\mathbf{E}(\boldsymbol{y}) = \frac{1}{2} \big \Vert \mathbf{S}\boldsymbol{x} - \mathbf{S}\boldsymbol{y} \big \Vert$
is nonconvex; therefore, its loss surface $\mathbf{E}$ may have local minimizers in addition to the global minimizers of the form $\boldsymbol{y}_\tau ^\ast (t) = \boldsymbol{x} (t-\tau)$ with $\vert \tau \vert < T$.
As a consequence, we formulate the problem of audio texture synthesis in loose terms of perceptual similarity.
We refer to \cite{schwarz2011dafx} for a literature review on texture synthesis, 
and to \cite{schwarz2016dafx} for a discussion of quantitative evaluation procedures.

Starting from a colored Gaussian noise $\boldsymbol{y}_0 (t)$ whose power spectral density matches $\mathbf{S}_1 \boldsymbol{x} (t,\lambda)$, we refine it by additive updates of the form $\boldsymbol{y}_{n+1}(t) = \boldsymbol{y}_n (t) + \boldsymbol{u}_n (t)$, where the term $\boldsymbol{u}_n (t)$ is defined recursively as $\boldsymbol{u}_n (t) = m \times \boldsymbol{u}_n (t) + \mu_n \boldsymbol{\nabla}\mathbf{E}(\boldsymbol{y}_n)(t)$.
In subsequent experiments, the momentum term is fixed at $m=0.9$ while the learning rate is initialized at $\mu_0 = 0.1$ and modified at every step according to a ``bold driver'' heuristic \cite{sutskever2013icml}.

\subsection{Gradient backpropagation in a scattering network}

Like deep neural networks, scattering networks consist of the composition of linear operators (wavelet transforms) and pointwise nonlinearities (complex modulus).
Consequently, the gradient $\mathbf{E}(\boldsymbol{y}_n)$ can be obtained by composing the Hermitian adjoints of these operator in the reverse order as in the direct scattering transform --- a method known as backpropagation \cite{bruna2013arxiv}.

First, we backpropagate the gradient of Euclidean loss for second-order scattering coefficients:
\begin{equation}
\boldsymbol{\nabla}\mathbf{U_2}\boldsymbol{y}(t,\lambda, \alpha, \beta) =
\Big((\mathbf{S_2}\boldsymbol{x}-\mathbf{S_2}\boldsymbol{y})
\overset{t}{\ast} \boldsymbol{\phi}\Big)
(t, \lambda, \alpha, \beta).
\end{equation}
Secondly, we backpropagate the second layer onto the first:
\begin{align}
\boldsymbol{\nabla}\mathbf{U_1}\boldsymbol{y}(t,\lambda) =
    \Big((\mathbf{S_1}\boldsymbol{x}-\mathbf{S_1}\boldsymbol{y})
\overset{t}{\ast} \boldsymbol{\phi}\Big)
(t, \lambda, \alpha, \beta) \nonumber \\ +
    \sum_{\alpha,\beta} \mathfrak{R}\Bigg( \Bigg[\dfrac{\big(\mathbf{U_1}\boldsymbol{y} \overset{t}{\ast} \boldsymbol{\bar{\psi}}_{\alpha} \overset{\log \lambda}{\ast} \boldsymbol{\bar{\psi}}_{\beta} \big)}{\big\vert\mathbf{U_1}\boldsymbol{y} \overset{t}{\ast} \boldsymbol{\bar{\psi}}_{\alpha} \overset{\log \lambda}{\ast} \boldsymbol{\bar{\psi}}_{\beta} \big\vert} \times \boldsymbol{\nabla}\mathbf{U_2}\boldsymbol{y} \Bigg]
    \nonumber \\
    \overset{t}{\ast} \boldsymbol{\psi}_{\alpha} \overset{\log \lambda}{\ast} \boldsymbol{\psi}_{\beta} \Bigg)(t, \lambda, \alpha, \beta),
\end{align}
where the symbol $\mathfrak{R}(z)$ denotes the real part of the complex number $z$.
Lastly, we backpropagate the first layer into the waveform domain:
\begin{equation}
    \boldsymbol{\nabla}\mathbf{E}(\boldsymbol{y})(t) =
    \sum_{\lambda} \mathfrak{R}\Bigg( \Bigg[ \dfrac{\boldsymbol{y}\overset{t}{\ast}\boldsymbol{\psi}_{\lambda}}{\vert\boldsymbol{y}\overset{t}{\ast}\boldsymbol{\psi}_{\lambda}\vert} \times \boldsymbol{\nabla}\mathbf{U_1}\boldsymbol{y} \Bigg] \overset{t}{\ast} \boldsymbol{\psi}_{\lambda}\Bigg)(t)
\end{equation}


Time--frequency scattering bears a strong resemblance with the set of spectrotemporal summary statistics developed by \cite{mcdermott2011neuron} to model the perception of auditory textures in the central auditory cortex.
A qualitative benchmark has shown that time--frequency scattering is advantaged if $\boldsymbol{x}(t)$ contains asymmetric patterns (\eg{} chirps), but that the two representations perform comparably otherwise \cite{lostanlen2017phd}.
Nevertheless, time--frequency scattering is considerably faster: the numerical optimizations of wavelet transforms and the recursive structure of backpropagation allows time--frequency scattering (both forward and backward) to be on par with real time on a personal computer, \ie{} about $20$ times faster than the other implementation.
Therefore, an audio snippet of a few seconds can be fully re-synthesized in less than a minute, which makes it relatively convenient for composing sound serendipitously.

\subsection{Creation: \emph{FAVN} (2016) and other pluriphonic pieces}
\label{sub:favn}

\emph{FAVN} is an electroacoustic piece that evokes issues surrounding late 19th-century psychophysics as well as Debussy's \emph{Pr\'elude \`a l'apr\`es-midi d'un faune} (1894), which itself is a musical adaption of St\'ephane Mallarm\'e's \emph{L'apr\`es-midi d'un faune} (1876).
To create \emph{FAVN}, we began by composing $47$ blocks of sound of duration equal to $21$ seconds, and spatialized across three audio channels.
These blocks are not directly created with time--frequency scattering; rather, they originate from the tools of the electroacoustic studio, such as oscillators and modulators.
After digitizing these blocks, we analyze them and re-synthesize them by means of time--frequency scattering.
We follow the algorithm described above: once initialized with an instance of Gaussian noise, the reconstruction is iteratively updated by gradient descent with a bold driver learning rate policy.
We stop the algorithm after $50$ iterations.

During the concert, the performer begins by playing the first iteration of the first block, and progressively moves forward in the reproduction of the piece, both in terms of compositional time (blocks) and computational time (iterations).
The Alte Oper Frankfurt, Frankfurt am Main, Germany premiered \emph{FAVN} on October 5th, 2016.
The piece was presented again at \emph{Geometry of Now} in Moscow in February 2017, and became a two-month exhibition at the Kunsthalle in Wien in November 2017, with a dedicated retrospective catalogue \cite{hecker2018book}.

In the liner notes of \emph{FAVN}, librettist Robin Mackay elucidates the crucial role of analysis-synthesis in the encoding of musical timbre:
\begin{quotation}
The analysis of timbre --- a catch-all term referring to those aspects of the \emph{thisness} of a sound that escape rudimentary parameters such a pitch and duration --- is an active field of research today, with multiple methods proposed for classification and comparison.
In \emph{FAVN}, Hecker effectively reverses these analytical strategies devised for timbral description, using them to synthesize new sonic elements.
In the first movement, a scattering transform with wavelets is used to produce an almost featureless ground from which an identifiable signal emerges as the texture is iteratively reprocessed to approximate its timbre. [Wavelets] correspond to nothing that can be heard in isolation, becoming perceptible only when assembled en masse. \cite{mackay2016favn}
\end{quotation}

We refer to \cite{lostanlen2019chapter} for further discussions on the musical implications of time--frequency scattering, and to \cite{hecker2009urbanomic,hecker2013book} on the musical aesthetic of Florian Hecker.
Since the creation of \emph{FAVN}, we have used time--frequency scattering to create four new pieces.

\emph{Modulator (Scattering Transform)} is a remix of Hecker's electronic piece \emph{Modulator} (2012), obtained by retaining the $50^{\textrm{th}}$ iteration of the gradient descent algorithm.
Editions Mego has released this remix in a stereo-cassette format.
In addition, we presented an extended version of the piece in a 14-channel format at the exhibition ``Florian Hecker - Formulations'' at the Museum f\"ur Moderne Kunst Frankfurt, Frankfurt am Main, Germany from November 2016 to February 2017.
In this multichannel version, $14$ loudspeakers in the same room play a different iteration number of the reconstruction algorithm.

\emph{Experimental Palimpsest} is an eight-channel variation upon \emph{Palimpsest} (2004), Hecker's collaboration with the Japanese artist Yasunao Tone, obtained by the same procedure.
This piece was premiered at the Lausanne Underground Film Festival, Lausanne, Switzerland, in October 2016.

\emph{Inspection (Maida Vale Project)} is a seven-channel piece for synthetic voice and computer-generated sound, performed live at BBC's Maida Vale studios in London and has been broadcast on BBC Radio 3 in December 2016, marking the BBC's first ever live binaural broadcast. An extended version, \emph{Inspection II}, will be released as a CD by Editions Mego, Vienna and Urbanomic, Falmouth, UK in Fall 2019.

\subsection{\emph{XAllegroX (scattering.m sequence)}}

Lastly, \emph{XAllegroX (scattering.m sequence)} is the remix of a dance music track by Lorenzo Senni, entitled \emph{XAllegroX} and originally released by Warp Record in Senni's \emph{The Shape of Trance to Come} LP (WAP406).
Like in Hecker's experimental pieces, we remixed \emph{XAllegroX} by, first, isolating a few one-bar loops from the original audio material, and secondly, by reconstructing them from their scattering coefficients.
While, at the first iteration, the loop sounds hazy and static --- or, a electronic musicians would call it, \emph{droney} --- it regains some of its original rhythmic content in subsequent iterations, thus producing a feeling of sonic rise that is fitting to the typical structuration of dance music.
The peculiarity of this \emph{scattering.m sequence} remix is that the musical ``rise'' is not produced over the well-known sonic attributes of frequency and amplitude (as is usually the case in electronic dance music), but on a relatively novel, joint parameter of texture; that is, a notion of sonic complexity which consists of the organization of frequencies and amplitude through time.
Therefore, the development of gradient backpropagation for time--frequency scattering over new avenues for musical creation with digital audio effects: in addition to remixing amplitude (by EQ-ing) and frequency (by phase vocoder), it is now possible to \emph{remix texture itself}, independently of amplitude and frequency.
Along the same lines of amplitude modulation (AM) and frequency modulation (FM), we propose to call this new musical transformation a meta-modulation (MM), because it operates over spectrotemporal modulations rather than on the direct acoustic content.
Future work will strive to further the understanding of MM, both from mathematical and compositional standpoints.

In July 2018, Warp released \emph{XAllegroX (scattering.m sequence)} as part of a remix 12" named \emph{The Shape of RemiXXXes to Come} (WAP425), hence the title of the present paper.
This remix has been pressed on vinyl and made available on all major digital music platforms.
The remix was aired on the Camarilha dos Quatro weekly podcast.
Mary Anne Hobbs, an English DJ and music journalist, shared another of the album songs on her BBC show ``6 Music Recommends''.
FACT listed the record as one of the must-haves of the month.

\section{Scale-rate digital audio effects}
\label{sec:scale-rate-dafx}
In this section, we introduce an algorithm to manipulate the finest time scales of spectrotemporal modulations (from $\SI{10}{\milli\second}$ to $\SI{1}{\second}$) while preserving both the temporal envelope and spectral envelope at a coarser scale (beyond $\SI{1}{\second}$).
As an example, we implement chirp rate reversal, a new digital audio effect that flips the pitch contour of every note in a melody, without need for tracking partials. This concept will be featured throughout in the pluriphonic sound piece \emph{Syn 21845} (2019), a sequel to Hecker's \emph{Statistique Synth\'etique aux \'epaules de cascades} (2019).

\subsection{Mid-level time scales in music perception}
The invention of digital audio technologies allowed composers to apply so-called \emph{intimate transformations} \cite{risset1999chapter} to music signals, affecting certain time scales of sound perception while preserving others.
The most prominent of such transformations is perhaps the phase vocoder \cite{roebel2010dafx}, which transposes melodies and/or warps them in time independently.
By setting $T$ to $\SI{50}{\milli\second}$, a wavelet-based phase vocoder disentangles frequencies belonging to the hearing range (above $\SI{20}{\Hz}$) from modulation rates that are afferent to the perception of musical time (below $\SI{20}{\Hz}$) \cite{kronland1988cmj}.
Frequency transposition is then formulated in $\mathbf{S}_1 \boldsymbol{x}$ as a translation in $\log_2 \lambda$ whereas time stretching is formulated as a homothety in $t$.

In its simplest flavor, the phase vocoder suffers from artifacts near transient regions: because all time scales beyond $T$ are warped in the same fashion, slowing down the tempo of a melody comes at the cost of a smeared temporal profile for each note.
This well-known issue, which motivated the development of specific methods for transient detection and preservation \cite{roebel2003dafx}, illustrates the importance of mid-level time scales in music perception, longer than a physical pseudo-period yet shorter than the time span between adjacent onsets \cite{leveau2008taslp}.

The situation is different in a time--frequency scattering network: the amplitude modulations caused by sound transients are encoded in the scale-rate plane $(\alpha, \beta)$ of spectrotemporal receptive fields \cite{anden2015mlsp}.
Therefore, time--frequency scattering appears as a convenient framework to address the preservation of such mid-level time scales in conjunction with a change in rhythmic parameters (meter and tempo); or, conversely, changes in articulation in conjunction with a preservation of the sequentiality in musical events.

\subsection{General formulation}

We propose to call \emph{scale-rate DAFx} the class of audio transformations whose control parameters are foremostly expressed in the domain $(t, \lambda, \alpha, \beta)$ of time--frequency scattering coefficients, and subsequently backscattered to the time domain by solving an optimization problem of the form
\begin{equation}
\boldsymbol{y}^\ast = 
\arg \min_y \big \Vert
f(\mathbf{S})\boldsymbol{x} -
\mathbf{S}\boldsymbol{y}
\big \Vert_2^2,
\end{equation}
where the functional $f(\mathbf{S})=(f_1(\mathbf{S}_1), f_2(\mathbf{S}_2))$ is defined by the composer.
Compared to Section \ref{sub:phase-retrieval}, the loss function in the equation above is not only nonconvex, but also devoid of a trivial global minimizer.
Indeed, if the image of the reproducing kernel Hilbert space (RKHS) associated to Equation \ref{eq:U2} by the complex modulus operator and low-pass filter $\boldsymbol{\phi}_T(t)$ (Equation \ref{eq:S2}) does not contain the function $f(\mathbf{S}\boldsymbol{x})$, then there is no constant-$Q$ transform $\mathbf{U}_1^* (t, \log_2 \lambda)$ whose smoothed STRF is $f_2(\mathbf{S}_2)$ ; and \emph{a fortiori} no waveform $\boldsymbol{y}^* (t)$ such that $\mathbf{S} \boldsymbol{y} = f(\mathbf{S})\boldsymbol{x}$.
In order to allow for more flexibility in the set of valid choices of $f$, we replace the definition of $\mathbf{S}_2 \boldsymbol{x}$ in Equation \ref{eq:S2} by
\begin{equation}
\mathbf{S}_2 \boldsymbol{x}(t, \lambda, \alpha, \beta) =
\big(
\mathbf{U}_2 \boldsymbol{x} \overset{t}{\ast} \boldsymbol{\phi}_T \overset{\log_2 \lambda}{\ast} \boldsymbol{\phi}_F
\big)(t, \lambda, \alpha, \beta),
\end{equation}
that is, a blurring over both time and frequency dimensions; and likewise for $\mathbf{S}_1 \boldsymbol{x}$.
This new definition guarantees that $\mathbf{S} \boldsymbol{x}$ is invariant to frequency transposition up to intervals of size $F$ (expressed in octaves), a property that is often desirable in audio classification \cite{anden2014tsp}.
Transposition-sensitive scattering (Equations \ref{eq:S1} and \ref{eq:S2}) are a particular case of transposition-invariant scattering (equation above) at the $F \rightarrow 0$ limit, \ie{} the Gaussian $\boldsymbol{\phi}_F$ becoming a Dirac delta distribution.

A thorough survey of scale-rate DAFx is beyond the scope of this article; in the sequel, we merely give some preliminary insights regarding their capabilities and limitations as well as a proof of concept.
With $Q \gg 12$ wavelets per octave in the constant-$Q$ transform and $F$ of the order of one semitone, scale-rate DAFx would fall within the well-studied application domain of vibrato transformations  \cite{roebel2011dafx}:
a translation of the variable $\log_2 \alpha$ (\resp $\log_2 \vert \beta \vert$) would cause a multiplicative change in vibrato rate (\resp depth).
Perhaps more interestingly, with $Q \ll 12$ and $F$ of the order of an octave, scale-rate DAFx address the lesser-studied problem of roughness transformations in complex sounds:
since the scattering transform captures pairwise interferences between pure tones within an interval of $Q^{-1}$ octaves or less \cite{anden2012dafx}, a translation of the variable $(\log_2 \lambda + \log_2 \alpha)$ would transpose the sound while preserving its roughness, whereas a translation of the variable $(\log_2 \lambda - \log_2 \alpha)$ would affect roughness while preserving the spectral centroid.
We believe that the capabilities of such transformations, both from computational and musical perspectives, are deserving of further inquiry.

\subsection{Example: controlling the axis of time with chirp inversion}
Because both Morlet wavelets $\boldsymbol{\psi}_\alpha (t)$ and $\boldsymbol{\psi}_\beta (\log_2 \lambda)$ have a symmetric profile, we have the following identity between Kronecker tensor products:
\begin{equation}
\boldsymbol{\psi}_\alpha \otimes \boldsymbol{\psi}_{-\beta} = \overline{\boldsymbol{\psi}_{-\alpha} \otimes \boldsymbol{\psi}_{\beta}}.
\end{equation}
Since the constant-$Q$ transform modulus $\mathbf{U}_1 \boldsymbol{x}$ is real-valued, the above implies that $\mathbf{S}_2 \boldsymbol{x}(t, \lambda, \alpha, -\beta) = \mathbf{S}_2 \boldsymbol{x}(t, \lambda, - \alpha, \beta)$.
In other words, flipping the sign of the modulation scale $\beta$ is equivalent to reversing the time axis in the wavelet $\boldsymbol{\psi}_\alpha$; or, again equivalently, to reversing the time axis in the constant-$Q$ transform $\mathbf{U}_1 \boldsymbol{x}$ around the center of symmetry $t$ before analyzing it with $\boldsymbol{\psi}_\alpha$ and $\boldsymbol{\psi}_\beta$.
From these observations, we define a chirp inversion functional $f(\mathbf{S}) = (f_1(\mathbf{S}_1), f_2(\mathbf{S}_2))$ where $f_1(\mathbf{S}_1) = \mathbf{S_1}$ and $f_2$ is parameterized as
\begin{equation}
\begin{split}
f_2:\mathbf{S}_2(t,\lambda,\alpha,\beta) \longmapsto &
\hphantom{+}\; \dfrac{1 + \boldsymbol{\sigma}(t)}{2} \times \mathbf{S}_2(t,\lambda,\alpha,\beta) \\
& + \dfrac{1 - \boldsymbol{\sigma}(t)}{2}  \times \mathbf{S}_2(t,\lambda,\alpha,-\beta),
\label{eq:chirp-inversion-functional}
\end{split}
\end{equation}
with $\boldsymbol{\sigma}(t)$ a slowly varying function at the typical time scale $T$.
Observe that setting $\boldsymbol{\sigma}(t) = 1$ leaves $\mathbf{S}$ unchanged;
that $\boldsymbol{\sigma}(t) = -1$ resembles short-time time reversal (STTR) of $\boldsymbol{x}(t)$ with half-overlapping windows of duration $T$ \cite{kim2014dafx}; 
and that $\boldsymbol{\sigma}(t) = 0$ produces a re-synthesized sound that is stationary, yet not necessarily Gaussian, up to the time scale $T$.
It thus appears that the parameter $\boldsymbol{\sigma}(t)$ in Equation \ref{eq:chirp-inversion-functional} is amenable to an ``axis of time'' knob that can be varied continuously through time within the range $[-1;\,1]$.

\begin{figure*}
\centering
   \begin{subfigure}[b]{\textwidth}
   \includegraphics[width=\textwidth]{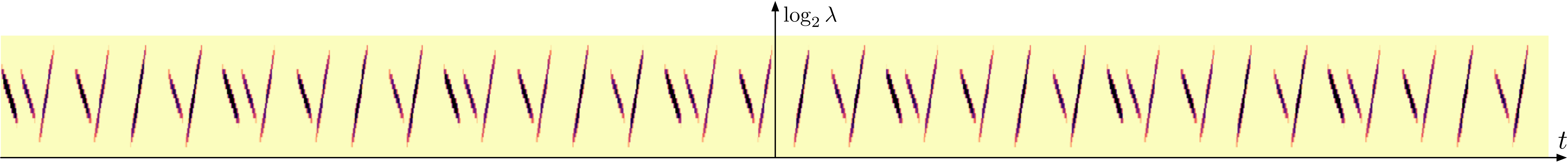}
   \caption{An interpretation of Steve Reich's \emph{Clapping Music} (1972) with synthetic chirps of varying rates, scales, and amplitudes.}
   \label{fig:chirp-rate-inversion-a} 
\end{subfigure}
\begin{subfigure}[b]{\textwidth}
   \includegraphics[width=\textwidth]{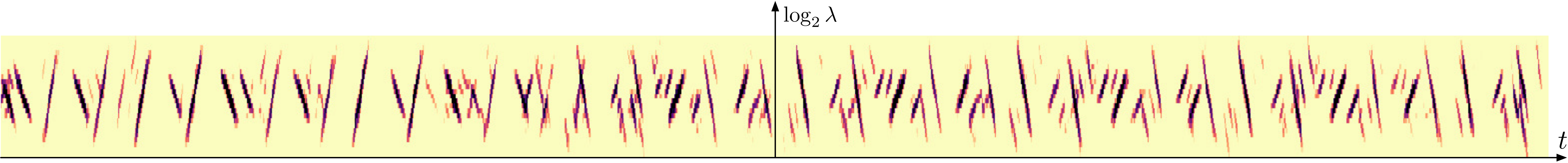}
   \caption{Re-synthesis after chirp rate inversion. The arrow of time goes forward for $t < 0$ and locally backward for $t > 0$.}
   \label{fig:chirp-rate-inversion-b}
\end{subfigure}
\caption{An example of chirp rate inversion with time--frequency scattering. Top: original audio material. Bottom: computer-generated output after $100$ iterations of gradient descent on time--frequency scattering coefficients. The chirp inversion functional follows a sigmoidal dynamic, as described in Equations \ref{eq:chirp-inversion-functional} and \ref{eq:sigma}. See Section 4 for details.}
\label{fig:chirp-rate-inversion}
\end{figure*}

As a proof of concept, Figure \ref{fig:chirp-rate-inversion-a} shows the constant-$Q$ transform of a repetitive sequence of synthetic chirps with varying amplitudes, frequential extents, and orientations; as well as its transformation by the functional $f$ described above, with
\begin{equation}
\boldsymbol{\sigma}(t) = \dfrac{1 - \exp\big(\frac{t}{ \tau}\big)}{1 + \exp\big(\frac{t}{\tau}\big)}
\label{eq:sigma}
\end{equation}
the sigmoid function with a time constant $\tau \gg T$. The frequency transposition invariance $F$ is set to $1$ octave.
We observe that, while the metrical structure of the original excerpt is recognizable at all times, the pitch contour of every musical event is identical to the original for $t \ll - \tau$ and inverted with respect to the original for $t \gg \tau$.
For $\vert t \vert < \tau$, there is a progressive metamorphosis between the ``forward time'' and ``backward time'' regimes.
The effect obtained in Figure \ref{fig:chirp-rate-inversion-b}, although relatively simple to express in the scale-rate domain, would be difficult to implement in the time--frequency domain.

\subsection{Towards digital audio effects on the pitch spiral}
One evident drawback of scale-rate DAFx is the need to manually adjust the frequency transposition invariance $F$ according to the analysis-synthesis task at hand.
Forgoing this calibration step would certainly streamline the creative process.
Furthermore, setting $F$ to any value above $1$ octave does not only affect spectrotemporal modulations but also the spectral envelope of $\mathbf{U}_1 \boldsymbol{x}(t,\lambda)$.
In the context of speech transformations, this undesirable phenomenon has led to the development of specific improvements to the phase vocoder \cite{roebel2010dafx}.

With the aim of addressing both of these shortcomings, we suggest replacing the resort to STRF in Equation \ref{eq:U2} $\mathbf{U}_2 \boldsymbol{x}(t,\lambda)$ by spiral scattering \cite{lostanlen2015dafx}, a convolutional operator cascading wavelet transforms along time, along log-frequencies, and across neighboring octaves.
Denoting by $\lfloor \log_2 \lambda \rfloor$ the octave index associated to the frequency $\lambda$ -- that is, the integer part of its binary logarithm -- the spiral scattering transform of $\boldsymbol{x}(t)$ writes as
\begin{multline}
\mathbf{U}_2 \boldsymbol{x}(t, \lambda, \alpha, \beta, \gamma) = \\
\Big \vert \mathbf{U}_1 \boldsymbol{x} \overset{t}{\ast} \boldsymbol{\psi}_{\alpha} \overset{\log_2 \lambda}{\ast}  \boldsymbol{\psi}_{\beta}
\overset{\lfloor \log_2 \lambda \rfloor}{\ast}  \boldsymbol{\psi}_{\gamma} \Big \vert  (t, \lambda)
\end{multline}
where $\boldsymbol{\psi}_\gamma$ is a Morlet wavelet of quality factor $Q = 1$ and center frequency $\gamma$, with $\vert \gamma \vert < \frac{1}{2}$ measured in cycles per octave.
Since spiral scattering disentangles temporal modulations of the nonstationary source-filter model \cite{lostanlen2017phd}, it is conceivable that source modulations and filter modulations could be manipulated independently in the space of spiral scattering coefficients.
In particular, the aforementioned effect of ``chirp rate reversal'' could be generalized to the modulations of the source-filter model.
For nonstationary harmonic tones, this would result in a reversal of melodic profile with preservation of the formantic profile, or vice versa.
Although the present article does not give a demonstration of such effects, it is worth remarking that their future implementation in the \emph{scattering.m} library would rely on the same principles as the gradient backpropagation of time--frequency scattering coefficients.

The procedure of rolling up the log-frequency axis into a spiral which makes a full turn at every octave, thus aligning power-of-two harmonics onto the same radius, is central to the construction of auditory paradoxes of pitch perception \cite{deutsch2008jasa} and has recently been applied to musical instrument classification \cite{lostanlen2017phd} and real-time pitch tuning \cite{helie2017dafx}; yet, to the best of our knowledge, never as a mid-level representation for DAFx.
As such, the theoretical framework between scale-rate DAFx and spiral DAFx lies at the interaction between two concurrent approaches in the DAFx community: sinusoidal modeling \cite{roebel2003dafx} and texture modeling via neural networks \cite{caracalla2016atiam}.
The former is more physically interpretable requires no training data, yet makes strong assumptions on the detectability of partials in the input spectrum; conversely, the latter is bereft of partial tracking, yet requires a training set and allows for less post hoc manipulations.
The long-term goal of scale-rate DAFx is to borrow from both of these successful approaches, and ultimately achieve a satisfying compromise between interpretability and flexbility in texture synthesis.

\section{Conclusion}

The past decade has witnessed a breakthrough of deep convolutional architectures for signal classification, with some noteworthy applications in speech, music, and ecoacoustics.
Yet there is, to this day, virtually no adoption of any recent deep learning system by electroacoustic music composers.
This is due to several shortcomings of deep learning in its current state, among which:
its high computational cost \cite{kim2015icml};
the need for a large dataset of musical samples, often supplemented with univocal human annotation \cite{engel2017icml};
the difficulty of synthesizing audio without artifacts \cite{donahue2018iclr};
and a certain opacity in the structure of the learned representation \cite{lample2017nips}.

In this article, we have argued that time--frequency scattering -- a deep convolutional operator involing little or no learning -- is adequate for several use cases of contemporary music creation.
We have supported our argumentation by three mathematical properties, which are rarely guaranteed in deep learning:
energy conservation (Section \ref{sec:time--frequency-scattering});
well-conditioned adjoint operators in closed form (Section \ref{sec:audio-texture-synthesis});
and psychophysiological interpretation in terms of modulation rates and scales (Section \ref{sec:scale-rate-dafx})..

All of the numerical applications presented here might, after enough effort, be implemented without resorting to time--frequency scattering at all.
Yet, one noteworthy trait of time--frequency scattering resides in its versatility: it connects various topics of DAFx that are seemingly disparate, such as coding \cite{velasco2011dafx}, texture synthesis \cite{schwarz2011dafx}, adaptive transformations \cite{roebel2003dafx}, and similarity retrieval \cite{pohle2005dafx}.
The guiding thread between these topics is that -- adopting the categories of Iannis Xenakis \cite{xenakis1989pnm} -- time--frequency scattering extracts musical information at the \emph{meso-scale} of musical motifs, allowing to put it in relation with both the \emph{micro-scale} of musical timbre and the \emph{macro-scale} of musical structures \cite{hecker2009urbanomic}.

From a compositional perspective, the appeal of time--frequency scattering stems from the possibility to generate sound with a holistic approach, devoid of intermediate procedures for parametric estimation; yet leaving some room to serendipity and surprise in the listening experience.
The best evidence of this fruitful trade-off between flexibility and interpretability is found in the breadth of computer music pieces that have resulted from the ongoing interaction between the two authors of this paper.
The earliest musical creation that was based on time--frequency scattering (\eg{} \emph{FAVN} in 2016) was conceived as an operatic experience with pluriphonic spatialization at the Alte Oper Frankfurt, Frankfurt am Main, Germany.
In contrast, \emph{Modulator (Scattering Transform)} (2017) is a stereo-cassette mix for Editions Mego; \emph{Inspection (Maida Vale Project)} (2017) is a live radio performance at the BBC; and lastly, \emph{XAllegroX (Hecker Scattering.m Sequence)} (2018) is the remix of a dance music track for Warp Records.

More than a fortuitous affinity of personal initiatives, the research-creation agenda that is outlined in the present paper wishes to espouse the noble tradition of \emph{musical research} \cite{risset1985esprit}, \ie{} a kind of creative process in which the conventional division of labor between scientists and artists is tentatively called into question.
Here, for the sake of crafting new music, a signal processing researcher (VL) becomes \emph{de facto} a computer music designer, while a composer (FH) takes on a role akin to a principal investigator \cite{cont2010icmc}.

\section{Acknowledgments}
This work is supported by the ERC InvariantClass grant 320959 and NSF awards 1633259 and 1633206.
The two authors wish to thank Bob Sturm for putting them in contact with each other; Lorenzo Senni for accepting that his record title, \emph{The Shape of RemiXXXes to Come}, is being reused as the title of the present article; and the anonymous reviewers of both DAFX-18 and DAFX-19 for their helpful comments.
%

\nocite{*}

\bibliographystyle{IEEEbib}
\bibliography{lostanlen_dafx2018.bib}

\end{document}